\documentclass[aps,10pt,twocolumn,showpacs,pra,citeautoscript,amsmath,amssymb,floatfix,nofootinbib,superscriptaddress,longbibliography]{revtex4-2}
\usepackage[left=2cm,right=2cm,top=2cm,bottom=2cm]{geometry}

\usepackage[colorlinks]{hyperref}
\hypersetup{
    allcolors  = {teal!70!black},
}

\usepackage[utf8]{inputenc}
\usepackage[english]{babel}
\usepackage[T1]{fontenc}
\usepackage{amsfonts}
\usepackage{dsfont}
\usepackage{bbm}
\usepackage{xfrac}

\usepackage{soul}

\usepackage{braket}
\usepackage{graphicx}

\usepackage{cleveref}
\usepackage{xcolor}

\usepackage{bbold}
\usepackage{amsthm}

\newcommand{\ulmshort}{Institute of Theoretical Physics, Ulm University, Ulm, Germany} 
\newcommand{\iqstshort}{Center for Integrated Quantum Science and Technology (IQST), 89081 Ulm, Germany} 
 
\begin{document}

\title{Bargmann Scenarios}

\author{Rafael Wagner}
\email{rafael.wagner@uni-ulm.de}
\affiliation{\ulmshort}
\affiliation{\iqstshort}

\date{\today}

\begin{abstract}
Considerable effort has been devoted to developing techniques for witnessing and characterizing quantum resources that emerge from collective properties of a set of states. In this context, Bargmann invariants play a central role: they witness coherence and related  resources, and underpin important applications. In this work, we introduce a unified formalism that fully characterizes and organizes the capability of Bargmann invariants to witness different manifestations of coherence in sets of states. It is formulated around the construction of \emph{Bargmann scenarios}, which specify relevant tuples of Bargmann invariants, and \emph{Bargmann polytopes}, which bound the values that said invariants can have when the states are incoherent. We study their basic geometry, connect them to existing formalisms, and illustrate their physical relevance. Our construction opens new opportunities for the certification of quantum devices and lays the path toward a full quantum resource theory based entirely on multivariate traces of states.
\end{abstract}

\maketitle

{\color{teal!70!black}\textit{Introduction.---}}Bargmann invariants~\cite{bargmann1964note} are multivariate traces of quantum states. For an arbitrary $n$-tuple of states $\vec\rho = (\rho_1,\ldots,\rho_n)$, the associated invariant is defined by
\begin{equation}
    \Delta(\vec\rho\,) := \mathrm{Tr}(\rho_1 \cdots \rho_n).
\end{equation}
They have recently attracted considerable attention~\cite{zhang2026surveybargmanninvariantsgeometric} due to their broad physical relevance, as they are connected to a wide variety of theoretical constructions and experimental settings, including Kirkwood--Dirac quasiprobability representations~\cite{kirkwood1933quantum,dirac1945analogy,wagner2024quantumcircuits,arvidssonshukur2024properties,schmid2024kirkwood,liu2025boundarykirkwooddiracquasiprobability,debievre2021complete}, out-of-time-order correlators~\cite{yunger2018quasiprobability,gonzalez2019out}, multi-time correlation functions~\cite{pedernales2014efficient}, weak values~\cite{wagner2023anomalous,hofmann2012complex,chiribella2024dimension}, noisy continuous sensing~\cite{yang2026quantumcramer}, indefinite causal order~\cite{ban2021sequential,gao2023measuring,azado2025measuringunitaryinvariantsquantum}, multiphoton indistinguishability~\cite{menssen2017distinguishability,jones2020multiparticle,minke2021characterizing,pont2022quantifying,rodari2024experimentalobservationcounterintuitivefeatures,rodari2024semideviceindependentcharacterizationmultiphoton,seron2023boson,giordani2021witnesses,giordani2020experimental,jones2023distinguishability,brod2019witnessing,annoni2025incoherentbehaviorpartiallydistinguishable}, overlap uncertainty relations~\cite{bong2018strong}, quantum thermodynamics~\cite{gherardini2024quasiprobabilities,lostaglio2022kirkwood,levy2020quasiprobability,hernandez2024projective,santini2023work,donati2024energetics,upadhyaya2024nonabelian},  quantum communication~\cite{elliott2025strictadvantagecomplexquantum}, and geometric phases~\cite{samuel1988general,akhilesh2020geometric,mukunda2003Wigner,mukunda2003Bargmann,rabei1999bargmann,hetnyi2026generating}. Moreover, they also clarify the role of nonclassicality in these constructions, as they can be used to characterize basis-independent quantum resources in collections of quantum states, including entanglement~\cite{zhang2024local,foulds2021controlledSWAP,cai2021entanglement,yu2021absolutely}, coherence~\cite{galvao2020quantum,designolle2021set}, contextuality~\cite{wagner2024coherence}, dimension~\cite{galvao2020quantum,giordani2021witnesses,wagner2024inequalities,giordani2023experimental}, imaginarity~\cite{fernandes2024unitary,miyazaki2022imaginarityfree}, non-Gaussianity~\cite{xu2025bargmanninvariantsgaussianstates}, and nonstabilizerness~\cite{wagner2024certifying,zamora2025semi}. 

Bargmann invariants also have practical appeal: they can be estimated directly with quantum circuits~\cite{quek2024multivariatetrace,oszmaniec2024measuring,simonov2025estimationmultivariatetracesstates,leifer2004measuring,shin2024rankneedestimatingtrace,faehrmann2025intheshadow}, and even more naturally on photonic platforms~\cite{pont2022quantifying,jones2020multiparticle,jones2023distinguishability,novo2026nativelinearopticalprotocolefficient,fujii2003exchange,foulds2024generalising,santosjunior2026coherence}. This makes them useful tools for device characterization and benchmarking~\cite{giordani2021witnesses,giordani2023experimental,giordani2020experimental}. Moreover, for a finite collection of states, sufficiently many Bargmann invariants capture \emph{all} relational properties~\cite{wagner2024quantumcircuits} (i.e. properties formulated in terms of unitary-invariant functions) of the set, since they generate the ring of unitary-invariant polynomials~\cite{oszmaniec2024measuring,wigderson2019mathematics}. As a result, whenever a numerical task depends only on such invariants, Bargmann invariants can provide a more economical representation and thereby reduce the computational resources required in practice~\cite{yang2026quantumcramer,martinezcifuentes2026calculatingtracedistancesbosonic}.

\begin{figure}[t]
    \centering
    \includegraphics[width=\columnwidth]{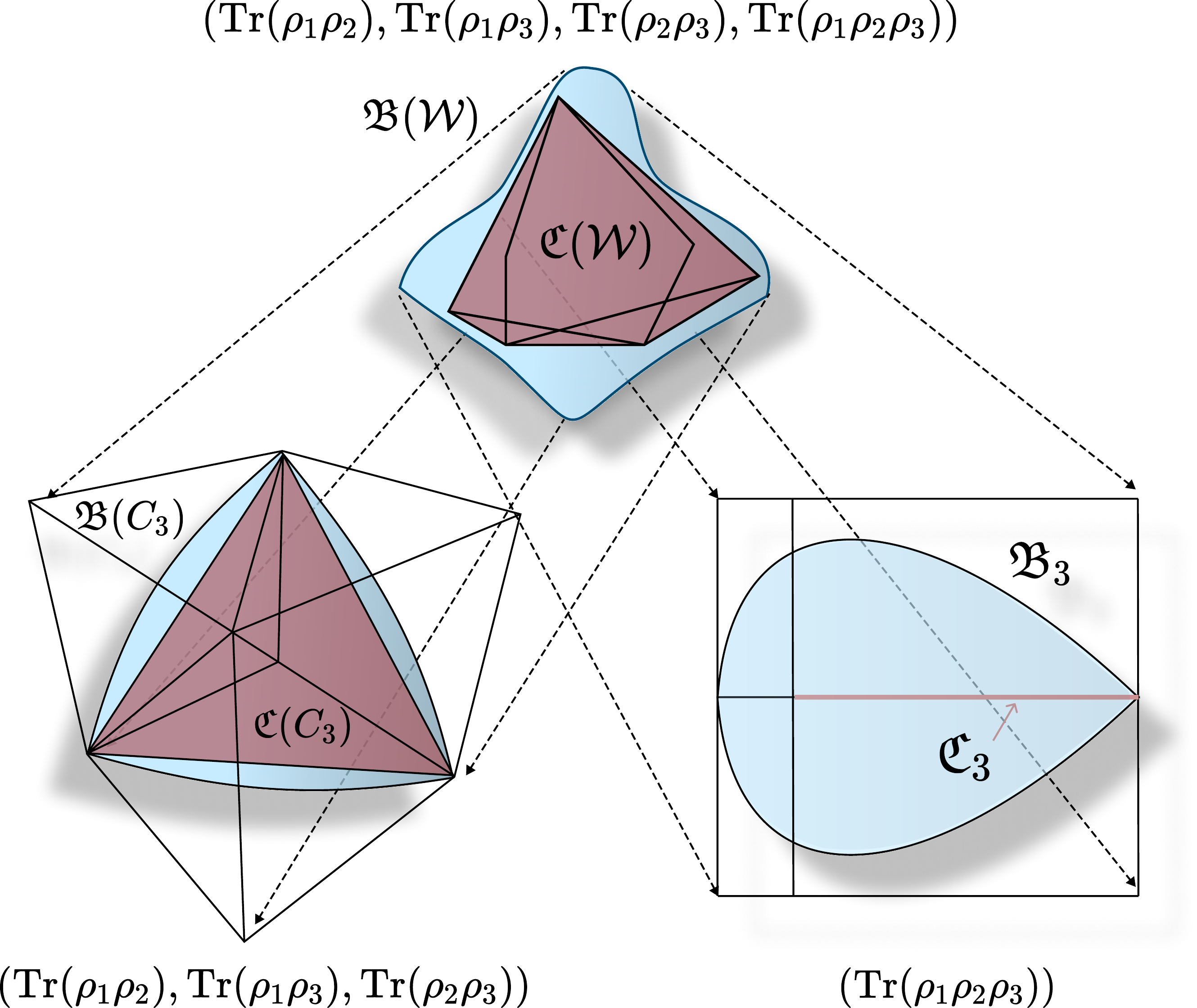}
    \caption{\textbf{Illustration of the main contribution.} Bargmann scenarios and their associated polytopes unify and generalize several previously studied constructions. The figure shows the quantum set $\mathfrak{B}(\mathcal{W})$ and the classical set $\mathfrak{C}(\mathcal{W})$ for $\mathcal{W}=\{(1,2),(1,3),(2,3),(1,2,3)\}$ first investigated in Refs.~\cite{eggeling2001separability,oszmaniec2024measuring}. The orthogonal projection onto binary words (left) reproduces the sets of Refs.~\cite{galvao2020quantum,giordani2021witnesses,bong2018strong}, while the projection onto three-letter words (right) yields the set studied in Refs.~\cite{fernandes2024unitary,xu2025numericalrangesbargmanninvariants,li2025bargmann,zhang2025geometrysets,pratapsi2025elementarycharacterizationbargmanninvariants}.}
    \label{fig:illustration_idea}
\end{figure}

Because Bargmann invariants play a central role in a wide range of contexts, the geometry of the full set of Bargmann invariants encodes rich information about nonclassicality and the relevant experimental constraints in many situations of interest. For instance, the unitary invariance of tuples of states $\vec \rho$ is closely related to the local-unitary (LU) invariance of multipartite states, where the geometry of the corresponding set provides information about separability conditions, as studied in Ref.~\cite{eggeling2001separability}. In the single-system setting, Refs.~\cite{wagner2024inequalities,galvao2020quantum} introduced an inequality-based framework for two-state overlaps and showed that they can witness basis-independent coherence of a set of states~\cite{designolle2021set}, a feature that was later applied experimentally to benchmark optical quantum devices~\cite{giordani2020experimental,giordani2021witnesses,giordani2023experimental,brod2019witnessing}. More recently, Ref.~\cite{fernandes2024unitary} showed that the complex values of Bargmann invariants themselves can witness basis-independent imaginarity of a set of states~\cite{miyazaki2022imaginarityfree}. Building on the fact that, for pure states, the positive semidefiniteness of the associated Gram matrix completely characterizes the quantum set~\cite{chefles2004physicaltransformations}, the sets of values that a Bargmann invariant can take have been fully characterized~\cite{fernandes2024unitary,xu2025numericalrangesbargmanninvariants,li2025bargmann,zhang2025geometrysets,pratapsi2025elementarycharacterizationbargmanninvariants}, followed by intriguing applications in photonics~\cite{rodari2024experimentalobservationcounterintuitivefeatures,rodari2025semi,hoch2025optimaldistillationphotonicindistinguishability}.

In this work, we introduce a unified formalism that fully characterizes and organizes the capability of Bargmann invariants to witness different manifestations of basis‑independent coherence in sets of quantum states (see Fig.~\ref{fig:illustration_idea}). The framework is built around \emph{Bargmann scenarios}, which specify the relevant quantum set of correlations formed by tuples of Bargmann invariants, and \emph{Bargmann polytopes}, which bound the classical subset of invariants given by incoherent states. We study the basic geometry of the quantum and classical sets, connect them to existing constructions (such as event‑graph polytopes~\cite{wagner2024inequalities,wagner2025coherenceandcontextuality} and the sets studied for imaginarity~\cite{fernandes2024unitary,xu2025numericalrangesbargmanninvariants,li2025bargmann,zhang2025geometrysets,pratapsi2025elementarycharacterizationbargmanninvariants}), and illustrate their physical relevance.

We show that some Bargmann polytopes can fully characterize the coherence of a set of states---that is, a tuple of states is set‑incoherent if and only if its vector of Bargmann invariants lies inside the corresponding polytope. This provides a convex treatment of the non‑convex set‑coherence problem from Ref.~\cite{designolle2021set} and unifies several earlier witnessing approaches based on overlaps, third‑order invariants, and higher‑order constraints. More broadly, our construction opens new opportunities for the certification of quantum devices, the classification of all possible ways in which sets of states can manifest basis-independent coherence, and lays the path toward a full quantum resource theory based entirely on Bargmann invariants.

{\color{teal!70!black}\textit{Bargmann scenarios.---}}We call a Bargmann scenario the specification of a set of labels identifying the relevant invariants we wish to study. To provide an instructive example for comparison, in a Bell scenario~\cite{brunner2014bell} one specifies labels of inputs and outputs for different parties, which in turn identify the correlations of interest. Here, by contrast, we consider labels such as $(1,2)$ and $(1,2,3)$, which identify the corresponding Bargmann invariants of interest, in this case $\mathrm{Tr}(\rho_1\rho_2)$ and $\mathrm{Tr}(\rho_1\rho_2\rho_3)$, respectively. 

Let $\mathcal{W} \subseteq \mathbbm{N}^*$ be a finite subset of finite sequences over the natural numbers, which we call \emph{words}. \footnote{Equivalently, $\mathcal{W}$ is a finite subset of the free monoid on $\mathbbm{N}$, denoted $\mathbbm{N}^*$, whose elements are finite sequences.} Define
\begin{equation}\label{eq:letters}
L := \bigcup_{\vec{w}\in\mathcal{W}} \{\, \ell \mid \ell \in \vec{w} \,\}
\end{equation}
as the set of all natural numbers appearing in the words of $\mathcal{W}$, which we refer to as \emph{letters}. The set $L$ specifies the labels of all states relevant to the scenario, while $\mathcal{W}$ specifies which relational quantities are included. The specification of the set $\mathcal{W}$ defines the Bargmann scenario.

We say that $\vec z \in \mathbbm{C}^\mathcal{W}$ is \emph{quantum realizable}~\cite{fernandes2024unitary,fraser2023estimationtheoreticapproachquantum} in a Hilbert space $\mathcal{H}$ if there exists a family of (possibly unnormalized) states $\vec\rho = (\rho_{\ell})_{\ell \in L}$ such that 
\begin{equation}\label{eq:quantum_realizable_points}
   z_{\vec w} = \text{Tr}\left( \rho_{\ell_1}\cdots \rho_{\ell_m}\right) = \Delta(\rho_{\vec w})
\end{equation}
for every word $\vec{w} = (\ell_1,\dots,\ell_m) \in \mathcal{W}$. Above, we have denoted $\rho_{\vec w} \equiv (\rho_{\ell_1},\ldots,\rho_{\ell_m})$ the finite sequence of states induced by $\vec w$. Tuples of Bargmann invariants are denoted as $\vec \Delta (\vec \rho\, ) \equiv (\Delta (\rho_{\vec w}))_{\vec w \in \mathcal{W}}$. The number $m$ is said to be the \emph{order} of the invariant $\text{Tr}\left( \rho_{\ell_1}\cdots \rho_{\ell_m}\right)$. 

Two words $\vec w_1$ and $ \vec w_2$ are \emph{cyclic conjugates} of each other if one can be obtained from the other by applying a finite number of cyclic permutations. Because the trace is cyclic, quantum realizations as described above yield the same value for all words that are equivalent under cyclic permutations. For example, $\Delta(\rho_{1,2,3}) = \Delta(\rho_{3,1,2})$ for every quantum realization since $(1,2,3)$ and $(3,1,2)$ are cyclically equivalent. Hence, we assume that the set $\mathcal{W}$ does not contain such redundant words.

Given a Bargmann scenario $\mathcal{W}$, the set of all conceivable correlations is simply $\mathbbm{C}^{\mathcal{W}}$. Among these, the quantum realizable ones form a subset $\mathfrak{Q}(\mathcal{W}) \subseteq \mathbbm{C}^{\mathcal{W}}$. Crucially, $\mathfrak{Q}(\mathcal{W})$ allows \emph{unnormalized} quantum realizations: the only requirement on the operators $\rho_\ell$ is that they be positive, i.e., $\rho_\ell \in \mathcal{B}(\mathcal{H})^+ = \{\rho \in \mathcal{B}(\mathcal{H}) \mid \mathrm{spec}(\rho) \subseteq [0,\infty)\}$ for some Hilbert space $\mathcal{H}$ and every $\ell \in L$. This is \emph{not} the usual quantum set of Bargmann invariants found in the literature. We denote the latter by $\mathfrak{B}(\mathcal{W})$, and it is defined by the same Eq.~\eqref{eq:quantum_realizable_points} but with the additional constraint that every $\rho_\ell$ be \emph{normalized}: $\rho_\ell \in \mathcal{D}(\mathcal{H}) = \{\rho \in \mathcal{B}(\mathcal{H})^+ \mid \operatorname{Tr}(\rho)=1\}$ for some $\mathcal{H}$ and all $\ell$.

This construction generalizes virtually all previously considered ones on sets of Bargmann invariants. To start, the description of the sets $\mathfrak{Q}(\mathcal{W})$ applies to realizations by any tuple of states, not only normalized ones as mostly considered in the frameworks of Refs.~\cite{galvao2020quantum,wagner2024inequalities,fernandes2024unitary}. These sets are relevant for investigating, for example, joint diagonalizability of POVMs, which can be viewed as collections of unnormalized states summing to the identity~\cite{designolle2021set}. 

Our framework also encompasses most constructions previously treated as special cases. The subsets of normalized realizations $\mathfrak{B}(\mathcal{W})$ recover the sets 
\begin{equation}
    \mathfrak{B}_n = \{\Delta (\vec \rho\,) \mid \vec\rho \in \mathcal{D}(\mathcal{H})^n\text{ for some }\mathcal{H}\},
\end{equation}
previously considered in Refs.~\cite{fernandes2024unitary,xu2025numericalrangesbargmanninvariants,li2025bargmann,zhang2025geometrysets,pratapsi2025elementarycharacterizationbargmanninvariants}, by choosing  $\mathcal{W} = \{(1,\ldots,n)\}$. As another example, interpreting the letters $L$ as nodes in a graph $G=(L,\mathcal{W}_E)$ (known as the \emph{event graph}~\cite{galvao2020quantum,wagner2024inequalities,wagner2025coherenceandcontextuality}), and letting the set $\mathcal{W}_E$ be given by the edges $\vec w = (\ell,\ell')$ in this graph, we recover the sets investigated by Refs.~\cite{galvao2020quantum,giordani2020experimental,giordani2021witnesses,giordani2023experimental,wagner2024certifying,wagner2024coherence,wagner2024inequalities,wagner2024quantumcircuits,brod2019witnessing,santosjunior2026coherence}.

Since our construction allows for words with repeated letters, it can describe quantum sets that were previously overlooked. For example, the Bargmann scenario $\left\{(1,1), (1,2), (2,2)\right\}$ yields the quantum set $\mathfrak{Q}(\mathcal{W})$ of all tuples of the form  $\vec \Delta(\vec \rho\,) =(\mathrm{Tr}(\rho_1^2), \mathrm{Tr}(\rho_2^2), \mathrm{Tr}(\rho_1\rho_2))$. These have complete information of the distance between the states $\rho_1$ and $\rho_2$ in $\vec{\rho}$~\cite{quek2024multivariatetrace} since 
\begin{equation*}
    d_2(\vec \Delta(\vec \rho \,)) = \text{Tr}(\rho_1^2)+\text{Tr}(\rho_2^2)-2\text{Tr}(\rho_1\rho_2) = \Vert \rho_1-\rho_2 \Vert_2^2
\end{equation*}
where $\Vert \cdot \Vert_2$ is the Schatten-2 norm.

As a last example, if we consider the Bargmann scenarios $\mathcal{W}_d \equiv \{(1),(1,1),\dots,(1,1,\ldots,1)\}$, where the last word has $d$ elements. These are of relevance for \emph{quantum spectroscopy}~\cite{johri2017entanglement,tirrito2024quantifying,wagner2024quantumcircuits,shin2024rankneedestimatingtrace,yirka2021qubitefficient,brun2004measuring,alves2003direct,ekert2002direct,horodecki2002method,zhang2025explicit}. In this context, the set of all Bargmann invariants is given by 
\begin{equation}
    \mathfrak{B}(\mathcal{W}_d) = \{(\mathrm{Tr}(\rho^k))_{k=1}^d \mid \rho \in \mathcal{D}(\mathcal{H}) \text{ for some }\mathcal{H}\},
\end{equation}
which carry the complete information of the spectrum of the quantum states. 

The scenarios above probe only partial relational information, tailored to a specific goal; at the opposite end, some Bargmann scenarios are complete. For any fixed integer $n\geq 1$ and space $\mathcal D(\mathbbm{C}^d)$, there is always a Bargmann scenario $\mathcal{W}^{(\rm full)}_{n,d}$ completely characterizing all the relational properties of $n$-tuples $\vec \rho \subseteq \mathcal D(\mathbbm{C}^d)$, in the sense that two tuples $\vec \rho$ and $\vec \sigma$ are unitarily equivalent iff they yield the same point $\vec \Delta (\vec \rho \,) = \vec \Delta (\vec \sigma \,)$ in $\mathfrak{Q}(\mathcal{W}_{n,d}^{(\rm full)})$. A non-minimal choice is provided by taking this Bargmann scenario to be the set of all labels encompassing all words of length $k=1,\ldots,d^2$ over letters $1,\ldots,n$~\cite{forbes2013explicit,oszmaniec2024measuring,wigderson2019mathematics,procesi1976invariant,razmyslov1974trace,formanek2006generating}.

{\color{teal!70!black}\textit{Geometry.}---}For arbitrary scenarios $\mathcal{W}$, the set $\mathfrak{Q}(\mathcal{W})$ has an intricate geometry; for instance, some scenarios yield non‑convex quantum sets. Take $\mathcal{W} = \{(1), (2), (1,2)\}$. While $\vec 0, \vec 1 \in \mathfrak{Q}(\mathcal{W})$, their convex combination $\frac12 \vec 0 + \frac12 \vec 1 \notin \mathfrak{Q}(\mathcal{W})$ because the first two components force a quantum realization to satisfy $\mathrm{Tr}(\rho_1)=\mathrm{Tr}(\rho_2)=\frac12$, which implies $\rho_1,\rho_2 \le \frac12 \mathbbm{1}$. The third component would then require $\frac12 = \mathrm{Tr}(\rho_1\rho_2) \le \mathrm{Tr}(\rho_1)\mathrm{Tr}(\rho_2) = \frac14$, a contradiction. Instead, if the scenario $\mathcal{W}$ is chosen such that every word has the same length, the set $\mathfrak{Q}(\mathcal{W})$ is a \emph{convex cone} (see App.~\ref{app:fixed-length-scenario}). If, additionally, we include in this scenario all words consisting of a single repeated letter $\{(\ell,\ldots,\ell)\mid \ell \in L\}$, the convex cone is also \emph{pointed} (see App.~\ref{app:fixed-length-scenario}). 

The subsets of normalized realizations $\mathfrak{B}(\mathcal{W})$ are not cones since $\mathfrak{B}(\mathcal{W}) \subseteq \{\vec z \in \mathbbm{C}^{\mathcal W} \mid \vert z_{\vec w}\vert \leq 1, \forall \vec w \in \mathcal{W}\}$ for every Bargmann scenario. Their geometry is significantly more challenging to analyze. In Appendix~\ref{app:fixed-length-scenario-normalized} we prove that a large class of Bargmann scenarios yield sets $\mathfrak{B}(\mathcal{W})$ which are \emph{star-shaped}~\cite{salazar2026quantumresourcetheoriesconvexity} with center at $\vec z = \vec 0$, generalizing an argument from Ref.~\cite{li2025bargmann}.

{\color{teal!70!black}\textit{Bargmann polytopes.---}}For every Bargmann scenario $\mathcal{W}$ there is a subset $\mathfrak{I}(\mathcal{W})$ of the quantum realizable points defined in the following manner. We say that  $ \vec z \in \mathfrak{I}(\mathcal{W})$ if there exists a family of normalized states $\vec \sigma = (\sigma_\ell)_{\ell \in L}$ in a Hilbert space $\mathcal{H}$, and an orthonormal  basis $\{\vert \lambda \rangle\}_{\lambda \in \Lambda}$ for which 
$\sigma_{\ell} = \sum_{\lambda \in \Lambda} p^{(\ell)}_\lambda \vert \lambda \rangle \langle \lambda \vert, \,\, \forall \ell \in L,$ 
such that 
\begin{align}\label{eq:incoherent_definition}
    z_{\vec w} &= \Delta(\sigma_{\vec w})= \sum_{\lambda\in \Lambda}\prod_{i=1}^m p^{(\ell_i)}_{\lambda} 
\end{align}
for every $\vec w = (\ell_1, \ldots, \ell_m) \in \mathcal{W}$~\cite[Sec. 5.1.1]{wagner2024quantumcircuits}. Equivalently, $\vec z \in \mathbbm{C}^{\mathcal{W}}$ is quantum realizable in $\mathcal{H}$ by some tuple of states $\vec \sigma$ incoherent~\cite{streltsov2017colloquium,baumgratz2014quantifying} relative to some basis $\{\ket \lambda\}_\lambda$. Note that we impose no constraint on the size of $\Lambda$. The \emph{Bargmann polytope}~\cite{ziegler1995lectures} $\mathfrak{C}(\mathcal{W})$, associated to a scenario $\mathcal{W}$, is given by the convex hull of $\mathfrak{I}(\mathcal{W})$. Succintly, $\mathfrak{C}(\mathcal{W}) = \mathrm{Conv}[\mathfrak{I}(\mathcal{W})]$.

It is simple to see that, if $\vec z \in \mathfrak{C}(\mathcal{W})$, then it is possible to interpret any  realization as follows: $\vec z \in \mathfrak{C}(\mathcal{W})$ if and only if there exists a finite alphabet $\Lambda$ and a family of jointly distributed random variables $(M_\ell)_{\ell\in L}$ taking values in $\Lambda$ such that, for every word $\vec w=(\ell_1,\ldots,\ell_m)\in\mathcal W$,
\begin{equation}\label{eq:probability_of_equals_main}
    z_{\ell_1,\ldots,\ell_m}
    =
    p(M_{\ell_1}=\cdots=M_{\ell_m}).
\end{equation}
In other words, each point of $\mathfrak{C}(\mathcal{W})$ is specified by a joint probability distribution $p$ on $\Lambda^L$ whose equal-output probabilities reproduce the coordinates of $\vec z$~\cite{wagner2024inequalities,wagner2025coherenceandcontextuality}.  

The operational relation between $\mathfrak{C}(\mathcal{W})$ and  $\mathfrak{I}(\mathcal{W})$ is provided by viewing the outputs of measurement results diagonalizing all states realizing points in $\mathfrak{I}(\mathcal{W})$ as jointly distributed random variables. Suppose that several parties labeled $\ell\in L$ receive their respective independent quantum states $\sigma_{\ell}$, and measure each state with the PVM $M$ given by $\{\vert \lambda \rangle \langle \lambda \vert \}_\lambda$. Then, for any word $\vec w$, the entry $z_{\vec w} = \Delta(\sigma_{\vec w})$ equals the probability that all parties listed in $\vec w$ obtain the same output. More formally, denoting by $M_\ell$ the random outputs of party $\ell$, there exists a (glob+al) probability distribution $p$ such that for every $\vec w \in \mathcal{W}$ it returns
\begin{equation}
    \Delta(\sigma_{\vec w}) = \sum_{\lambda \in \Lambda }\prod_{i=1}^m p(M_{\ell_i} = \lambda)= p(M_{\ell_1}= \cdots = M_{\ell_m}).
\end{equation}
This condition is an extension of the transitivity of equality condition used in Refs.~\cite{galvao2020quantum,wagner2024inequalities} to the case of Bargmann invariants of arbitrary order. It is simple to see that   $\mathfrak{C}(\mathcal{W})$ is a convex $0/1$-polytope~\cite{ziegler1995lectures}: it is a convex set whose extreme points are deterministic assignments which preserve the transitivity of equality (see App.~\ref{app:Bargmann_polytopes}). In particular, if we let $\mathcal{W}$ be given by $\mathcal{W}_E$ introduced earlier, we recover exactly the event-graph polytopes introduced in Refs.~\cite{galvao2020quantum,wagner2024inequalities}. {Therefore, operationally, we can view $\mathfrak{C}(\mathcal W)$ as the extension of $\mathfrak{I}(\mathcal W)$ obtained by allowing a source of shared randomness to mix deterministic incoherent realizations, each of which corresponds to a classical assignment of outputs to labels.~\footnote{{This is related to a standard convexification strategy in prepare-and-measure scenarios~\cite{brask2026quantumcorrelationsprepareandmeasurescenarios}. When shared randomness is allowed, the classical correlation set becomes convex and is in fact a polytope~\cite{gallego2010device}, without shared randomness, convexity is no longer guaranteed~\cite{bowles2014certifying,devicente2017shared}.}}}

This formulation generalizes the construction in Ref.~\cite[Appendix VI]{oszmaniec2024measuring}, while also taking inspiration from it. There, the Bargmann polytope for the scenario $\mathcal{W}=\{(1,2),(1,3),(2,3),(1,2,3)\}$ is explicitly characterized, described by the facet-defining inequalities
$\vec z= {\vec z}\,^*$, $z_{1,2,3}\ge 0$,
$z_{1,2,3}\le z_{\ell,\ell'}$ for $\ell\neq \ell'$,
and $z_{1,2,3}\ge \sfrac{1}{2}(z_{1,2}+z_{1,3}+z_{2,3}-1)$.

{\color{teal!70!black}\textit{Connection with set coherence.---}}Several works have shown that Bargmann invariants can witness the basis-independent notion of set coherence introduced by Designolle \emph{et al.}~\cite{designolle2021set}. We now show that our framework comprehensively generalizes these approaches: it provides a convexified framework to study set coherence, yields (in principle) necessary and sufficient conditions for it formulated entirely in terms of tuples of Bargmann invariants, and, more broadly, organizes the various manifestations of coherence in sets of states within a unified structure.

Recall that a tuple of states $\vec \rho=(\rho_\ell)_{\ell=1}^n \in \mathcal D(\mathcal H)^n$ is called \emph{set incoherent} if there exists a unitary $U$ and a family of diagonal density matrices $\vec \sigma=(\sigma_\ell)_{\ell=1}^n$, all diagonal in a common orthonormal basis $\{\ket{\lambda}\}_{\lambda\in\Lambda}$, such that $\sigma_\ell = U \rho_\ell U^\dagger$ for all $\ell \in L$. It follows immediately that, for any scenario $\mathcal W$, every point $\vec z \in \mathfrak{B}(\mathcal W)\setminus \mathfrak{C}(\mathcal W)$ witnesses set coherence for any tuple of normalized states $\vec \rho$ realizing it via $\vec z = \vec \Delta(\vec \rho)$. For full scenarios such as $\mathcal W_{n,d}$, membership in $\mathfrak{C}(\mathcal W_{n,d})$ provides  necessary and sufficient conditions for set coherence of $n$-tuples of $d$-dimensional normalized states. 

The construction of Bargmann polytopes is, in fact, richer than simply deciding whether a set of states exhibits coherence. Rather, the polytope $\mathfrak{C}(\mathcal W)$ organizes the different manifestations of set coherence through the violation of its defining inequalities---in a similar vein to how the local polytope organizes different manifestations of Bell nonlocality~\cite{brunner2014bell}. To provide a few examples, set imaginarity can be understood as the violation of the constraints $z_{\vec w}=z_{\vec w^*}$, where $\vec w^*$ denotes the word with reversed order~\cite{fernandes2024unitary}. Violations of other equality constraints~\cite{wagner2025coherenceandcontextuality,li2025multistateimaginaritycoherencequbit} of the form $z_{\vec w}=z_{\pi\vec w}$ associated with permutations $\pi$ of the word elements represent different manifestations of set coherence. Furthermore, the family of overlap inequalities introduced in Ref.~\cite{wagner2024inequalities}, which witness both coherence and Hilbert-space dimension~\cite{giordani2023experimental}, provides yet another distinct manifestation of coherence in collections of states.

Bargmann scenarios can also clarify the interplay between nonconvexity at the level of state tuples and convex geometry at the level of invariant tuples. While convex combinations of set incoherent tuples may yield set coherent ones~\cite{designolle2021set}, and also $\mathfrak{I}(\mathcal{W})$ may be, in general, a non-convex set (see App.~\ref{app:Bargmann_polytopes}), the associated Bargmann-invariant points necessarily leave the polytope $\mathfrak{C}(\mathcal{W})$ whenever  the quantum realizable points cannot be interpreted as given by jointly distributed random variables (or, alternatively, convex combinations of incoherently realizable tuples) satisfying the transitivity of equality via Eq.~\eqref{eq:probability_of_equals_main}. In our framework, the nonconvexity of set incoherence is thus resolved upon passing to the Bargmann invariant quantum realizability description.   

 \begin{figure}[t]
    \centering
    \includegraphics[width=0.47\textwidth]{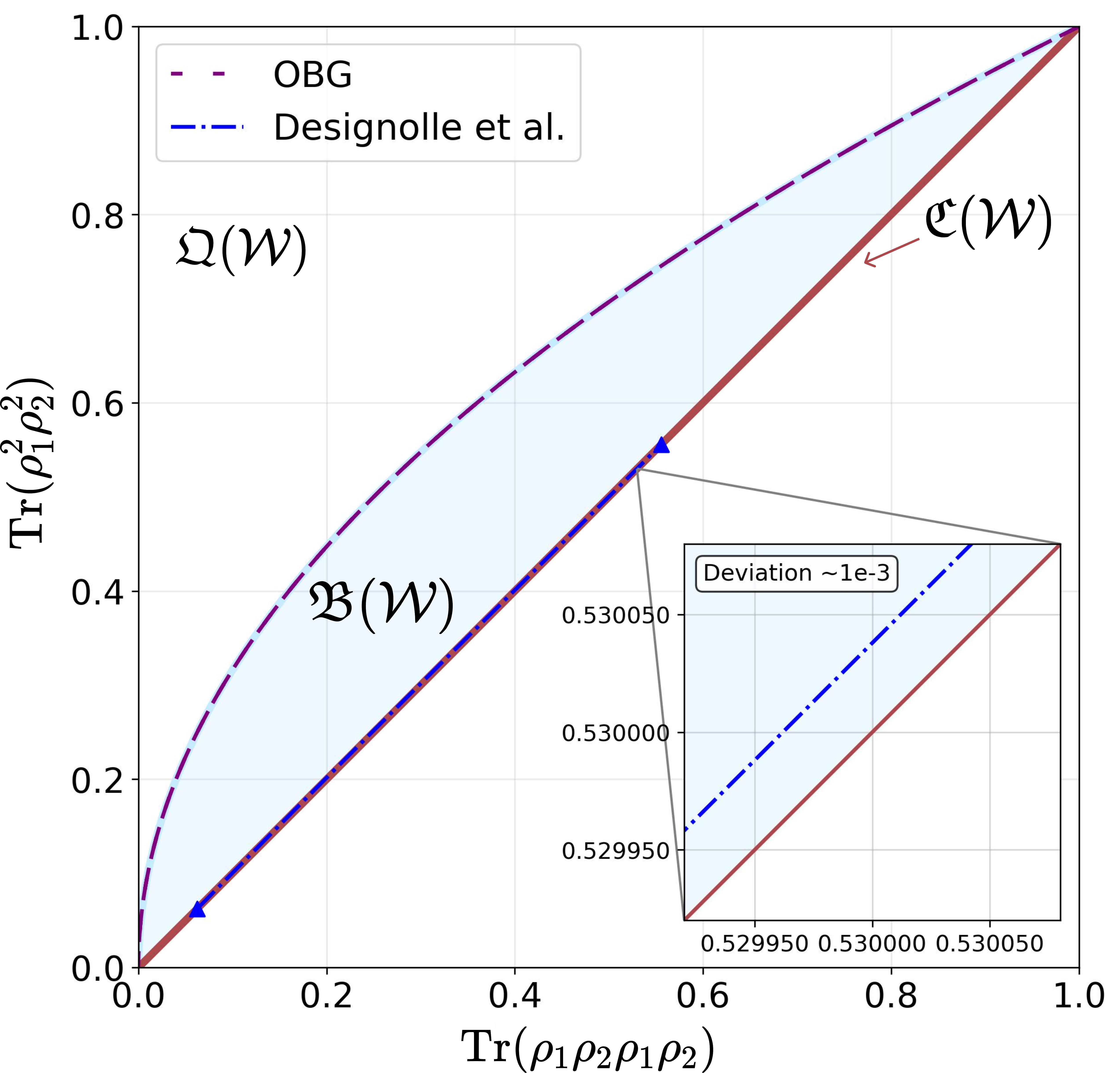}
    \caption{\textbf{Quantum and classical sets for  $\mathcal{W} = \{(1,1,2,2),(1,2,1,2)\}$.} $\mathfrak{C}(\mathcal{W})$ is the diagonal $x=y$ within $[0,1]$ (solid line), $\mathfrak{B}(\mathcal{W})$ is given by $0\le y^2\le x\le y\le 1$ (shaded region), and $\mathfrak{Q}(\mathcal{W})$ is the convex cone generated by $\mathfrak{B}(\mathcal{W})$. The OBG family of tuples from Eq.~\eqref{eq:obg_states} (dashed curve) realizes the boundary $x=y^2$. The family of tuples from Designolle \emph{et al.}~\cite{designolle2021set} from Eq.~\eqref{eq:designolle_states} realizes a curve (dash‑dotted) which lies inside $\mathfrak{B}(\mathcal{W})\setminus\mathfrak{C}(\mathcal{W})$ for $0<\omega<1$---thus witnessing set coherence---but lies in $\mathfrak{C}(\mathcal{W})$ for $\omega \in \{0,1\}$ (blue triangles).}
    \label{fig:simplest}
\end{figure}

Our proposed framework also opens new possibilities for coherence witnessing. In the two-state overlap framework of Refs.~\cite{wagner2024inequalities}, no inequality is known to be violated by quantum realizations of tuples $\vec \rho=(\rho_1,\rho_2)$. In our framework, by contrast, witnesses capable of capturing coherence of a pair of states are ubiquitous. Take as an example the scenario $\mathcal{W}$ with $L=\{1,2,3,4\}$ consisting of all non-cyclically equivalent words of lengths two, three, and four, for which one obtains the facet-defining inequality
\begin{equation}
    -z_{14}+z_{142}+z_{143}-z_{1423}\leq 0
\end{equation}
bounding $\mathfrak{C}(\mathcal{W})$. This inequality is violated by quantum realizations relative to tuples $\vec \rho=(\rho,\rho,\sigma,\sigma)$, yielding the operator relation
\begin{equation*}
    -\mathrm{Tr}(\rho\sigma)+\mathrm{Tr}(\rho^2\sigma)+\mathrm{Tr}(\rho\sigma^2)-\mathrm{Tr}(\rho\sigma\rho\sigma)\leq 0,
\end{equation*}
which is expressed in terms of second-, third-, and fourth-order Bargmann invariants. An example of a violation is provided by the paradigmatic example $\rho = \vert 0 \rangle \langle 0 \vert$ and $\sigma = \vert + \rangle \langle + \vert$, which attains the maximal violation of $0.25$. 


We can use the characterization of the two-word scenario $\mathcal W=\{(1,2,1,2),(1,1,2,2)\}$ as a fully worked-out example. In this case, the only facet-defining constraint is given by
$z_{1212}=z_{1122},$ 
which characterizes the polytope $\mathfrak{C}(\mathcal{W})=\{(x,y)\in[0,1]^2\mid x=y\}$.  The quantum set of normalized realizations is
\begin{equation}
\mathfrak{B}(\mathcal{W})=\{(x,y)\in\mathbbm{R}^2\mid 0\leq y^2\leq x\leq y\leq 1\},
\end{equation}
whose boundary is given by the intersection of the line $x=y$ and the parabola $x=y^2$, as shown in Fig.~\ref{fig:simplest} (see App.~\ref{app:worked_out_set}). The quantum set of unnormalized realizations is given by $\mathfrak{Q}(\mathcal{W}) = \{\lambda \vec z \mid \lambda \geq 0, \vec z \in \mathfrak{B}(\mathcal{W})\}$, i.e. the convex cone generated by $\mathfrak{B}(\mathcal{W})$.

Within $\mathfrak{B}(\mathcal{W})$, we identify two interesting families of points. First, the one generated by Oszmaniec--Brod--Galv\~ao (OBG) states~\cite{oszmaniec2024measuring,zhang2025geometrysets,li2025bargmann,pratapsi2025elementarycharacterizationbargmanninvariants}, defined for $n\geq 2$ by \begin{equation}\label{eq:obg_states}
\ket{\psi_k(\theta)}=\cos(\sfrac{\theta}{2})\ket 0+\sin(\sfrac{\theta}{2})e^{2\pi i k/n}\ket 1
\end{equation} 
with $k=0,1,\ldots,n-1$ and $\theta\in[0,2\pi)$. For  $n=2$, varying $\theta$ in the pairs $\vec \psi = \left(\psi_0(\theta), \psi_1(\theta) \right)$
traces the boundary of $\mathfrak{B}(\mathcal{W})\setminus \mathfrak{C}(\mathcal{W})$, as shown in 
Fig.~\ref{fig:simplest}. Second, following  Designolle \emph{et al.}~\cite{designolle2021set} (see also Ref.~\cite{li2025multistateimaginaritycoherencequbit}), we consider the tuples defined by the convex mixture 
\begin{equation}\label{eq:designolle_states}
\vec \xi=\omega \vec \rho+(1-\omega)\vec \sigma,
\end{equation}
with $0 \leq \omega \leq 1$, of the set incoherent pairs $ \vec \rho = 
\bigl(\vert 0 \rangle \langle 0 \vert,\sfrac{1}{3}\vert 0 \rangle \langle 0 \vert+\sfrac{2}{3}\vert 1 \rangle \langle 1 \vert\bigr)$
and $\vec \sigma = 
\bigl(\vert + \rangle \langle + \vert,\sfrac{1}{4}\vert + \rangle \langle + \vert+\sfrac{3}{4}\vert - \rangle \langle - \vert\bigr).$ Here, $\vert \pm  \rangle = \sfrac{1}{\sqrt{2}}(\vert 0 \rangle \pm \vert 1 \rangle)$.  
Varying $\omega$, we find---in complete accordance with Ref.~\cite{designolle2021set}---that for the deterministic values $\omega\in \{0,1\}$ (blue triangles in Fig.~\ref{fig:simplest}) the points lie inside the classical set, whilst the remainder of the curve lies outside, as shown in the inset of Fig.~\ref{fig:simplest}. 

{\color{teal!70!black}\textit{Discussion and outlook.---}}We have introduced Bargmann scenarios and Bargmann polytopes as a unified framework for organizing and characterizing the quantum and classical constraints on tuples of Bargmann invariants. Beyond recovering and generalizing several previously studied constructions, our approach provides a convex description of set coherence entirely in terms of relational data.

We anticipate that this framework will be particularly well suited for benchmarking set coherence and other relational properties in quantum devices. As a concrete example, Refs.~\cite{wang2026loworderbargmanninvariant,wagner2026commutativitysinglebargmanninvariant} have recently showed that the Bargmann scenario $\mathcal{W} = \{(1,2,1,2),(1,1,2,2)\}$ we have fully characterized  stands out as it completely distinguishes between coherent and incoherent realizations, i.e., any $\vec \rho = (\rho_1,\rho_2)$ in any finite-dimensional space $\mathcal{H}$ is set coherent \emph{iff} $\vec \Delta(\vec \rho\,) \notin \mathfrak{C}(\mathcal{W})$. Another promising avenue is analyzing nuances in partial multiphoton indistinguishability~\cite{menssen2017distinguishability,jones2023distinguishability,annoni2025incoherentbehaviorpartiallydistinguishable}, where Bargmann invariants already play a central role~\cite{brod2019witnessing,pont2022quantifying,shchesnovich2015partial,shchesnovich2018collective}. Future work should deepen our understanding of the geometry of the quantum sets $\mathfrak{B}(\mathcal{W})$ and $\mathfrak{Q}(\mathcal{W})$ beyond the simple cases analyzed here, and further investigate the structure of Bargmann polytopes $\mathfrak{C}(\mathcal{W})$, including their facet inequalities and their connections to other forms of coherence, such as nonstabilizerness. 

{\color{teal!70!black}\textit{Acknowledgments.---}}
We thank Ernesto Galv\~{a}o, Mao-Sheng Li, Lin Zhang, and Rui Soares Barbosa for discussions. We would like to especially thank Yan-Ling Wang for pointing out a subtle aspect of the relation between Bargmann polytopes and incoherent realizations. The author acknowledges support from the European Research Council (ERC) under the European Union's
Horizon 2020 research and innovation programme (Grant Agreement No. 856432, HyperQ) and from the Alexander von Humboldt Foundation.

\bibliography{Bibliography}

\begin{appendix}

\section{Convexity of quantum sets in fixed word-length Bargmann scenarios}\label{app:fixed-length-scenario}

Suppose that the Bargmann scenario $\mathcal{W}$ is length-homogeneous, meaning that all words $\vec w \in \mathcal{W}$ have equal length, i.e. every word $\vec w = (\ell_1, \ldots, \ell_m)$ for some integer $m\geq 1$. We want to show that $\mathfrak{Q}(\mathcal{W})$ is convex. Then, let $\vec z_1, \vec z_2 \in \mathfrak{Q}(\mathcal{W})$ be arbitrary points and let $\vec \rho, \vec \sigma$ be two sets of quantum states realizing them according to Eq.~\eqref{eq:quantum_realizable_points} in the main text. Specifically, for every $\vec w \in \mathcal{W}$ it holds that $({z_1})_{\vec w} = \Delta(\rho_{\vec w})$ and $(z_2)_{\vec w} = \Delta(\sigma_{\vec w})$. We want to show that for any $\alpha \in (0,1)$ the combination $\vec z = \alpha \vec z_1 + (1-\alpha)\vec z_2 \in \mathfrak{Q}(\mathcal{W})$. To see this, let   $a=\alpha^{1/m}$ and $b = (1-\alpha)^{1/m}$. Let us define $\vec\varrho$ via $\varrho_\ell = a\rho_\ell\oplus b \sigma_\ell$ for every $\ell \in L$. Note that for every such $\alpha$ the numbers $a,b \geq 0$ and thus the direct sum constitutes a positive operator. We now show that this tuple realizes $\vec z$. For every $\vec w = (\ell_1,\ldots,\ell_m) \in \mathcal{W}$ we have that 
        \begin{align*}
        \Delta(\varrho_{\vec w}) &= \text{Tr}\left(\prod_{i=1}^m \varrho_{\ell_i}\right) \\
        &= \text{Tr}\left(\prod_{i=1}^m (a\, \rho_{\ell_i} \oplus b\, \sigma_{\ell_i}) \right)  \\
        &=a^m \text{Tr}\left(\prod_{i=1}^m \rho_{\ell_i}\right)+b^m\text{Tr}\left(\prod_{i=1}^m \sigma_{\ell_i}\right)\\
        &= \alpha \Delta({{\rho}\,}_{\vec{w}}) + (1-\alpha)\Delta({{\sigma}\,}_{\vec{w}})\\
        &=\alpha (z_1)_{\vec w} + (1-\alpha)(z_2)_{\vec w}.
    \end{align*}
This concludes the proof. Note that for these scenarios we allow some words to be formed by repetitions of the same  letter. For example, we can have the scenario $\mathcal{W} = \{(1,1),(2,2),(1,2)\}$ considered in the main text. Note also that these sets are \emph{cones}. Note that $\vec 0 \in \mathfrak{Q}(\mathcal{W})$ for every scenario $\mathcal{W}$, and thus in particular in these sets. Moreover, if we have that $\vec z \in \mathfrak{Q}(\mathcal{W})$ it is clear that for all $\lambda \geq 0$ we have that $\lambda \vec z \in \mathfrak{Q}(\mathcal{W})$ since, given $\vec \rho$ realizing $\vec z$ we can choose $\vec \sigma = \lambda^{1/m}\vec \rho$ which will realize $\lambda \vec z$. 

We can also show that if the scenario $\mathcal{W}$ has all words $\{(\ell,\ldots,\ell) \mid \ell \in L\}$ (as the two letter example above) this convex cone is also \emph{pointed}. A cone $C$ is pointed if $C \cap -C = \{\,\vec 0\,\}$. Equivalently, whenever $\vec z \in C$ and $-\vec z \in C$ we have that $\vec z = \vec 0$.  Suppose that there is a linear functional $f$ such that for all $\vec z \in C$ it holds that $f(\vec z\,) \geq 0$. We can use said functional to analyze the pointedness of a cone $C$ since in this case, whenever we find that this functional satisfies that $f(\vec z\,) = 0$ iff $\vec z = \vec 0$, we can conclude that the cone $C$ is pointed. In our case, take $f$ given by $$f(\vec z\,) = \sum_{\ell \in L}z_{(\ell,\ldots,\ell)}.$$ This functional satisfies precisely the desired property. In this case, for every quantum realizable point $\vec z \in \mathfrak{Q}(\mathcal{W})$ we must have that 
\begin{equation}
    f(\vec z\,) = \sum_{\ell \in L}\mathrm{Tr}(\rho_\ell^m)
\end{equation}
where we have taken any representative $\vec \rho$ realizing $\vec z$. If $f(\vec z\,) = 0$ is zero, then for all $\ell$ we must have that $\rho_\ell = \mathbb{0}$ since, for all $\rho_\ell$ we have $\mathrm{Tr}(\rho_\ell^m) = \sum_{\lambda_\ell \in \mathrm{spec}(\rho_\ell)}\lambda_\ell^m$ which is zero only if all elements $\lambda_\ell = 0$, implying that $\rho_\ell = \mathbb{0}$. We conclude that $\mathfrak{Q}(\mathcal{W}) \cap -\mathfrak{Q}(\mathcal{W}) = \{\,\vec 0\,\}$.

\section{Star-shaped property of quantum sets realized by normalized states}\label{app:fixed-length-scenario-normalized}

Let us now consider length-homogeneous Bargmann scenarios $\mathcal{W}$ where every word contains at least two different letters. First, note that in all such scenarios the quantum set of tuples of Bargmann invariants realizable by normalized states contains the zero vector $\vec 0 \in \mathfrak{B}(\mathcal{W})$. To show that these sets have a star-shaped geometry, we simply need to show that for every $\nu \in [0,1]$ we have that $\vec z \in \mathfrak{B}(\mathcal{W})$ implies that $\nu \vec z \in \mathfrak{B}(\mathcal{W})$. In particular, suppose that $\vec \rho \in \mathcal{D}(\mathcal{H})^L$ realizes $\vec{z}$ in $\mathcal{H}$. Then, let $\vec \varrho$ be given by 
\begin{equation}
    \varrho_\ell = \nu^{1/m} \rho_\ell \oplus (1-\nu^{1/m})\vert \ell \rangle \langle \ell \vert 
\end{equation}
where $\vert \ell \rangle$ denote vectors of the canonical orthonormal basis for $\mathbbm{C}^{|L|}$. In this case, for all words $\vec w = (\ell_1, \ldots, \ell_m) \in \mathcal{W}$ we have that 
\begin{align*}
    \Delta (\varrho_{\vec w}) &= \mathrm{Tr}\left(\prod_{i=1}^m \nu^{1/m}\rho_{\ell_i} \oplus (1-\nu^{1/m})\vert \ell_i \rangle \langle \ell_i \vert\right)\\
    &=\nu \mathrm{Tr}\left(\prod_{i=1}^m \rho_{\ell_i}\right) + (1-\nu^{1/m})^m\mathrm{Tr}\left( \prod_{i=1}^m \vert \ell_i \rangle \langle \ell_i \vert\right)\\
    &=\nu \mathrm{Tr}\left(\prod_{i=1}^m \rho_{\ell_i}\right) = \nu \Delta(\rho_{\vec w}).
\end{align*}
Above, we have used that in every word there are at least two distinct letters, implying that the multivariate trace on the basis states must equal zero.

Denote by $\odot$ element-wise multiplication of matrices (also known as the Hadamard product). As a side remark, we also note that $$\vec z_1,\vec z_2 \in \mathfrak{B}(\mathcal{W}) \implies \vec z_1 \odot \vec z_2 \in \mathfrak{B}(\mathcal{W}),$$ for every Bargmann scenario $\mathcal{W}$ because if $\vec \rho$ realizes $\vec z_1$ in $\mathcal{H}$ and $\vec \sigma$ realizes $\vec z_2$ in $\mathcal{K}$ then $\vec \varrho $ defined by $\varrho_\ell = \rho_\ell \otimes \sigma_\ell$ for every $\ell \in L$ realizes $ \vec z_1 \odot \vec z_2$ in $\mathcal{H} \otimes \mathcal{K}$. This is so because for every $\vec w = (\ell_1,\ldots,\ell_m) \in \mathcal{W}$, we have that
\begin{align*}
    \mathrm{Tr}\left(\prod_{i=1}^m (\rho_{\ell_i} \otimes \sigma_{\ell_i})\right) &= \mathrm{Tr}\left(\prod_{i=1}^m \rho_{\ell_i} \otimes \prod_{i=1}^m \sigma_{\ell_i})\right) \\
    &=\mathrm{Tr}\left(\prod_{i=1}^m \rho_{\ell_i}\right)\mathrm{Tr}\left(\prod_{i=1}^m \sigma_{\ell_i}\right). 
\end{align*}

\section{Bargmann polytopes}\label{app:Bargmann_polytopes}

In this appendix we show that the sets $\mathfrak{C}(\mathcal{W})$ are convex 0/1-polytopes. This is a simple extension of the results of Ref.~\cite{wagner2024inequalities} and of the discussion in an appendix of Ref.~\cite{oszmaniec2024measuring}.

We begin by proving that $\mathfrak{C}(\mathcal{W})$ is convex. By definition, this follows from the fact that $\vec z \in \mathfrak{C}(\mathcal{W})$ if and only if there exists a finite alphabet $\Lambda$ and a family of jointly distributed random variables $(M_\ell)_{\ell\in L}$ taking values in $\Lambda$ such that, for every word $\vec w=(\ell_1,\ldots,\ell_m)\in\mathcal W$,
\begin{equation}\label{eq:probability_of_equals}
    z_{\ell_1,\ldots,\ell_m}
    =
    p(M_{\ell_1}=\cdots=M_{\ell_m}).
\end{equation}
In other words, each point of $\mathfrak{C}(\mathcal{W})$ is specified by a joint probability distribution $p$ on $\Lambda^L$ whose equal-output probabilities reproduce the coordinates of $\vec z$.

Now let $\alpha\in[0,1]$ and take arbitrary points $\vec x,\vec y\in\mathfrak{C}(\mathcal{W})$. Let $(M_\ell)_\ell$ and $(M'_\ell)_\ell$ be joint families of random variables realizing $\vec x$ and $\vec y$, respectively, with alphabets $\Lambda$ and $\Lambda'$, and joint distributions $p$ and $p'$. Define a new alphabet $\Lambda'' := \Lambda\,\sqcup\,\Lambda'$ and a joint distribution $q$ on $(\Lambda'')^L$ by setting
\begin{align}
    &q(\vec \lambda\,'')
=\begin{cases}
\alpha\, p(\vec \lambda),
& \text{if } \lambda_{\ell}''=(1,\lambda_{\ell})\ \text{for all } \ell,\\[1mm]
(1-\alpha)\, p'(\vec \lambda'),
& \text{if } \lambda_{\ell}''=(2,\lambda'_{\ell})\ \text{for all } \ell,\\[1mm]
0, & \text{otherwise}.
\end{cases}
\end{align}
Above, we use $\vec \lambda = (\lambda_\ell)_{\ell \in L}$, and similarly for $\vec \lambda'$ and $\vec \lambda''$. In words, $q$ is the convex mixture of $p$ and $p'$ supported on two disjoint copies of their alphabets. It is immediate that $q$ is normalized and nonnegative. Let $(M_\ell'')_{\ell\in L}$ be the corresponding jointly distributed random variables. Then, for every word $\vec w=(\ell_1,\ldots,\ell_m)\in\mathcal W$,
\begin{align*}
&q(M_{\ell_1}''=\cdots=M_{\ell_m}'')
=\\
&\alpha\, p(M_{\ell_1}=\cdots=M_{\ell_m})
+(1-\alpha)\, p'(M'_{\ell_1}=\cdots=M'_{\ell_m}) \\
&=
\alpha\, x_{\vec w}+(1-\alpha)\, y_{\vec w}.
\end{align*}
Hence $\alpha\vec x+(1-\alpha)\vec y\in\mathfrak{C}(\mathcal W)$, so $\mathfrak{C}(\mathcal W)$ is convex.

We now proceed to show that $\mathfrak{C}(\mathcal{W})$ is a convex subset of the convex hull of a finite set of deterministic 0/1 points (thus concluding that it is a convex 0/1 polytope). To start, given any Bargmann scenario $\mathcal{W}$, since the set of all letters $L$ is finite we can fix once and for all a finite alphabet $\Lambda = \{1,\ldots,|L|\}$ without loss of generality~\cite{wagner2024inequalities}. Then, for each assignment $\alpha: L \to \Lambda$---corresponding to an assignment of basis vectors $\vert \lambda \rangle \langle \lambda \vert $ to labels $\ell \in L$---we define the deterministic vectors using the indicator functions 
\begin{equation}
\lambda_\alpha(\ell_1,\ldots,\ell_m)=
\begin{cases}
1, & \text{if } \alpha(\ell_1)=\cdots=\alpha(\ell_m),\\
0, & \text{otherwise},
\end{cases}
\end{equation}
for every word $(\ell_1, \ldots,\ell_m) \in \mathcal{W}$. Note that we \emph{do not} assume that all words in the scenario have the same length. Let $\vec z\in\mathfrak{C}(\mathcal{W})$. By definition, there exist probability distributions $p^{(\ell)}=(p^{(\ell)}_\lambda)_{\lambda\in\Lambda}$, for all $\ell \in L$, 
such that for every word $\vec w=(\ell_1,\ldots,\ell_m)\in\mathcal{W}$, $$z_{\vec w}=\sum_{\lambda\in\Lambda}\prod_{i=1}^m p^{(\ell_i)}_\lambda.$$ Here, we are using the shorthand notation $p^{(\ell)}_\lambda \equiv p(M_\ell = \lambda)$. Let us now define, for each assignment $\alpha \in \Lambda^L$ the weight
\begin{equation}
q_\alpha:=\prod_{\ell\in L}p^{(\ell)}_{\alpha(\ell)}.
\end{equation}
These are nonnegative for all $\alpha$, and sum to one:
\begin{equation}
\sum_{\alpha\in\Lambda^L}q_\alpha
=
\sum_{\alpha \in \Lambda^L}\prod_{\ell\in L}p^{(\ell)}_{\alpha(\ell)} = \prod_{\ell\in L}\sum_{\lambda\in\Lambda}p^{(\ell)}_\lambda
=
1.
\end{equation}
Hence $\{q_\alpha\}_{\alpha\in\Lambda^L}$ is a family of convex weights. Denote $S_{\vec w} = \{\ell_1,\ldots,\ell_m\} \subseteq L$. Now fix an arbitrary word $\vec w=(\ell_1,\ldots,\ell_m)\in\mathcal{W}$. Then, by construction, 
\begin{align*}
\sum_{\alpha\in\Lambda^L} q_\alpha\,\lambda_\alpha(\vec w)
&=
\sum_{\alpha\in\Lambda^L}
\left(\prod_{\ell\in L} p^{(\ell)}_{\alpha(\ell)}\right)
\lambda_\alpha(\vec w)\\
&=
\sum_{\lambda\in\Lambda}\,\,
\sum_{\substack{\alpha\in\Lambda^L: \\\alpha(\ell_1)=\cdots=\alpha(\ell_m)=\lambda}}\,\,
\prod_{\ell\in L} p^{(\ell)}_{\alpha(\ell)} \\
&=
\sum_{\lambda\in\Lambda}
\left(\prod_{i=1}^m p^{(\ell_i)}_\lambda\right)
\sum_{\beta\in\Lambda^{L\setminus S_{\vec w}}}\,
\prod_{\ell\in L\setminus S_{\vec w}} p^{(\ell)}_{\beta(\ell)}\\
&=
\sum_{\lambda\in\Lambda}
\left(\prod_{i=1}^m p^{(\ell_i)}_\lambda\right)
\prod_{\ell\in L\setminus S_{\vec w}}
\Biggr(\underbrace{\sum_{\mu\in\Lambda} p^{(\ell)}_\mu}_{=1}\Biggr)\\
&=
\sum_{\lambda\in\Lambda}
\prod_{i=1}^m p^{(\ell_i)}_\lambda\\
&=
z_{\vec w}.
\end{align*}
Therefore $
\vec z\in \operatorname{conv}\{\lambda_\alpha:\alpha\in\Lambda^L\}.$ Above, the second equality comes from partitioning $\Lambda^L$ according to the common value taken by the letters appearing in the word $(\ell_1,\ldots,\ell_m)$:
\begin{equation}
    \Lambda^L = \bigcup_{\lambda \in \Lambda}\left\{\alpha \in \Lambda^L \mid \alpha(\ell_1)= \cdots = \alpha(\ell_m)= \lambda\right\}.
\end{equation}
In other words, for each $\lambda\in\Lambda$, we collect precisely those assignments $\alpha$ for which $\alpha(\ell_1)=\cdots=\alpha(\ell_m)=\lambda$. The third equality comes from separating the contribution of the letters appearing in $\vec w$ from the contribution of the remaining letters. For a fixed $\lambda$, the letters in the word are forced to take the value $\lambda$, so their contribution is $\prod_{i=1}^m p^{(\ell_i)}_\lambda$. The letters \emph{not} appearing in the word are unconstrained, and summing over all possible values of each such letter produces a factor $\sum_{\mu\in\Lambda}p^{(\ell)}_\mu=1$ for every $\ell\in L\setminus S_{\vec w}$. 

Moreover, each deterministic vector $\lambda_\alpha$ belongs to $\mathfrak{C}(\mathcal{W})$: simply choose, for every $\ell\in L$, the delta distribution $p^{(\ell)}_\lambda=\delta_{\lambda,\alpha(\ell)}.$ Then the defining formula reproduces $\lambda_\alpha$. Since $\mathfrak{C}(\mathcal{W})$ is convex by construction, it follows that
\begin{equation}
\mathfrak{C}(\mathcal{W})
=
\operatorname{conv}\{\lambda_\alpha:\alpha\in\Lambda^L\}.
\end{equation}
Since $\{\lambda_\alpha:\alpha\in\Lambda^L\}$ is finite, we conclude that $\mathfrak{C}(\mathcal{W})$ is a convex 0/1 polytope.

{Note that one could equivalently define Bargmann polytopes operationally as the set of Bargmann tuples realized by jointly distributed random variables with equal-output probabilities, and then recover the relation $\mathfrak{C}(\mathcal{W}) = \mathrm{Conv}[\mathfrak{I}(\mathcal{W})].$
In this sense, $\mathfrak C(\mathcal W)$ is the classical convex closure of the incoherent realizable tuples. The inclusion $\mathfrak I(\mathcal W)\subseteq \mathfrak C(\mathcal W)$ follows from the operational identification between incoherent diagonal states and classical random variables. The reverse inclusion is a consequence of the fact that $\mathfrak C(\mathcal W)$ is a convex $0/1$ polytope whose extreme points are deterministic incoherent assignments.

The sets $\mathfrak C(\mathcal W)$ and $\mathfrak I(\mathcal W)$ can be distinct whenever $\mathfrak I(\mathcal W)$ is non-convex. For instance, for $\mathcal{W} = \{(1,1), (1,2), (2,2)\}$, the pure incoherent pairs $(\rho_1,\rho_2)=(\ket0\bra0,\ket1\bra1)$ and $(\sigma_1,\sigma_2)=(\ket0\bra0,\ket0\bra0)$ yield the points  $\vec \Delta (\vec \rho \,) = (1,0,1)$ and $\vec \Delta (\vec \sigma\,) = (1,1,1)$. Their midpoint
\begin{equation}
    \vec z = \sfrac{1}{2} \, \vec \Delta (\vec \rho \,)+\sfrac{1}{2}\, \vec \Delta (\vec \sigma\,) = \left(1,\,\sfrac{1}{2}, \, 1 \right)
\end{equation}
belongs to $\mathfrak C(\mathcal W)$ and $\mathfrak{B}(\mathcal{W})$ but not to $\mathfrak I(\mathcal W)$. This illustrates how repeated-letter coordinates can induce non-convexity at the level of incoherent realizations for $\mathfrak{I}(\mathcal{W})$, thus objectively allowing in some situations for simpler coherence witnesses, while the corresponding classical Bargmann polytope remains convex.

To conclude, let us exemplify how to algorithmically find $\mathfrak{C}(\mathcal{W})$. To do so, we use a simple extension of the techniques from Refs.~\cite{galvao2020quantum,wagner2024inequalities}. Take, for example, the Bargmann scenario
\begin{align*}
    \mathcal{W} = \bigr\{ &(1,2,1,2), (1,1,2,2),\\ &(1,3,1,3),(1,1,3,3),\\ &(2,3,2,3),(2,2,3,3) \bigr \}.
\end{align*}
We start by considering all possible deterministic assignments for $\vec z$. In this case, we have $2^6$ deterministic assignments (vertices of $[0,1]^6$). However, not all  assignments preserve the transitivity of equality assumption, i.e. that entries $z_{\vec w}$ are given via Eq.~\eqref{eq:probability_of_equals} for some set of jointly distributed random variables $(M_\ell)_{\ell \in L}$. For example, take the point $(1,0,1,1,1,1)$. If it could be realized as a point in $\mathfrak{C}(\mathcal{W})$ it would imply both that $$z_{1,2,1,2} = p(M_{1} = M_2=M_1=M_2) = p(M_1=M_2) = 1$$ as well as $$z_{1,1,2,2} = p(M_1=M_1=M_2=M_2)=p(M_1=M_2) = 0,$$ a clear contradiction. The point $(1,1,0,0,1,1)$ also yields a contradiction since the first two entries imply that $M_1=M_2$ and the last two imply that $M_2=M_3$ while the middle ones imply $M_1 \neq M_3$. However, the point $(1,1,0,0,0,0)$ is allowed. Proceeding in this manner we see that the allowed points are: 
\begin{align*}
    \mathrm{ext}(\mathfrak{C}(\mathcal{W})) = \bigr \{ &(0,0,0,0,0,0), (1,1,1,1,1,1), \\&(1,1,0,0,0,0), (0,0,1,1,0,0), \\&(0,0,0,0,1,1)\bigr \}
\end{align*}
which define the V-representation~\cite{ziegler1995lectures} of the convex polytope $\mathfrak{C}(\mathcal{W})$~\cite{ziegler1995lectures}. We thus have a polytope which mirrors the simplest polytope from Ref.~\cite{galvao2020quantum}
\begin{equation}
    z_{1122}+z_{1133}-z_{2233} \leq 1
\end{equation}
and the other two sign permutations, together with the conditions $z_{\ell \ell \ell' \ell'}=z_{\ell \ell' \ell \ell'}$ for all $\ell, \ell' \in \{1,2,3\}$.
}

\section{Quantum set for the two-word scenario}\label{app:worked_out_set}

Given the Bargmann scenario $$\mathcal{W} = \left\{(1,1,2,2),(1,2,1,2)\right\},$$ in this appendix we show that the quantum set $\mathfrak{B}(\mathcal{W})$ of quantum realizable vectors $\vec z \in \mathbbm{C}^\mathcal{W}$ by Bargmann invariants of normalized tuples of states is given by
\begin{equation}\label{eq:quantum_set_proof_appendix}
    \mathfrak{B}(\mathcal{W})=\left\{(x,y)\in\mathbbm{R}^2\mid 0\leq y^2\leq x\leq y\leq 1\right\}.
\end{equation}
For arbitrary quantum states $\rho_1,\rho_2 \in \mathcal{D}(\mathbbm{C}^d)$, we start by choosing $U$ such that $U \rho_1U^\dagger = \sum_{i=1}^d\lambda_i \vert \lambda_i \rangle \langle \lambda_i \vert $ and let $\sigma = U \rho_2 U^\dagger$. In this case, 
\begin{align}
    \Delta(\rho_{1,2,1,2}) &= \sum_{i,j=1}^d\lambda_i\lambda_j \langle \lambda_i \vert \sigma \vert \lambda_j \rangle \langle \lambda_j \vert \sigma \vert \lambda_i \rangle \\
    &= \sum_{i,j=1}^d\lambda_i\lambda_j \vert \sigma_{ij}\vert^2
\end{align}
and
\begin{align}
    \Delta(\rho_{1,1,2,2}) &= \sum_{i,j=1}^d\lambda_i\lambda_j\delta_{ij}\langle \lambda_j \vert \sigma^2\vert \lambda_i \rangle \nonumber \\
    &= \sum_{i=1}^d\lambda_i^2\langle \lambda_i \vert \sigma^2 \vert \lambda_i \rangle \nonumber \\
    &=\sum_{i,j=1}^d\lambda_i^2 \langle \lambda_i \vert \sigma \vert \lambda_j \rangle \langle \lambda_j \vert \sigma \vert \lambda_i \rangle \nonumber \\
    &= \sum_{i,j=1}^d\lambda_i^2 \vert \sigma_{ij}\vert ^2,
\end{align}
where we have denoted $\sigma_{ij} := \langle \lambda_i \vert \sigma \vert \lambda_j\rangle$ the matrix elements of $\sigma$ relative to the eigenbasis of $\{\vert \lambda_i \rangle \langle \lambda_i \vert \}_i$. Now, we start by noting that letting $(x,y) = (\Delta(\rho_{1,2,1,2}), \Delta(\rho_{1,1,2,2}))$ we have 
\begin{align}
    y-x&= \sum_{i,j=1}^d (\lambda_i^2-\lambda_i\lambda_j)\vert \sigma_{ij}\vert^2\\
    &=\frac{1}{2}\sum_{i,j=1}^d(\lambda_i-\lambda_j)^2\vert \sigma_{ij}\vert^2 \geq 0
\end{align}
from which we conclude $x\leq y \leq 1$. Now, note that the matrix $M_{ij} = \vert \sigma_{ij}\vert^2$ is positive semidefinite,~\footnote{This follows from the fact that $M_{ij} = (\sigma \odot \sigma^*)_{ij}$ where $\sigma^*$ denotes complex conjugation of the density matrix $\sigma$ written in the eigenbasis of $\rho_1$. Since $\sigma$ is positive semidefinite, $\sigma^*$ is also, and by the Schur product theorem, $M$ is also positive semidefinite.} and therefore the Gram matrix for some  vector $\vec v$, so that we can write $M_{ij} = \langle v_i \vert v_j \rangle $. We can now use the Cauchy--Schwarz inequality
\begin{align}
    y^2 &= \left(\sum_{i,j}\lambda_i^2 \vert \sigma_{ij}\vert^2\right)^2 \leq \left(\sum_{i,j} \lambda_i\langle v_i \vert v_j \rangle\right)^2 \nonumber \\&= \left \vert \left \langle \sum_i \lambda_i v_i \Bigr \vert \sum_j v_j\right\rangle \right \vert ^2   \leq \,\Bigr\Vert \sum_i \lambda_i v_i \Bigr\Vert^2 \Bigr \Vert \sum_j v_j \Bigr \Vert^2\nonumber \\
    &=\left \langle \sum_i \lambda_i v_i \Bigr \vert \sum_j \lambda_j v_j\right \rangle \left \langle \sum_i  v_i \Bigr \vert \sum_j v_j\right \rangle \nonumber \\
    &=\sum_{ij}\lambda_i\lambda_j \vert \sigma_{ij}\vert^2 \sum_{ij}\vert \sigma_{ij}\vert^2\leq \sum_{ij}\lambda_i\lambda_j \vert \sigma_{ij}\vert^2 = x
\end{align}
from which we conclude $0 \leq y^2\leq x$. Above, in the first inequality we have used that since $0 \leq \lambda_i \leq 1$ then $\lambda_i^2\leq \lambda_i$ for all $i$. On the second line, we have used the Cauchy--Schwarz inequality, and on the last line we have used that 
\begin{equation}
    1 \geq \mathrm{Tr}(\sigma^2)= \sum_{i}\langle \lambda_i \vert \sigma^2 \vert \lambda_i \rangle = \sum_{ij}\vert \sigma_{ij}\vert^2.
\end{equation}

Given any quantum set $\mathfrak{B}(\mathcal{W})$ for a Bargmann scenario $\mathcal{W}$ we denote by $\mathfrak{B}^{(d)}(\mathcal{W})$ the set of all possible pairs realizable by $\vec \rho$ in a $d$-dimensional Hilbert space~\cite{wagner2025coherenceandcontextuality}. With this notion, it is also possible to show that for this two-word scenario it holds that $$\mathfrak{B}(\mathcal{W}) = \mathfrak{B}^{(2)}(\mathcal{W}),$$ i.e., that every point $(x,y)$ can be realized by some pair of single-qubit states. 

Given any pair of normalized states $\rho_1,\rho_2 \in \mathcal{D}(\mathbbm{C}^2)$, let us choose $U$ such that $U \rho_1 U^\dagger = \sfrac{1}{2}(\mathbbm{1}+r\,X)$ where $X$ is the first Pauli matrix, $0 \leq r \leq 1$, and also so that $\rho_2$ is entirely described by 
\begin{equation}
    U\rho_2U^\dagger = \frac{1}{2}\left(\mathbbm{1}+s \cos(\theta) X+s\sin(\theta)Z \right)
\end{equation}
with $0\leq s \leq 1$. Due to unitary invariance this does not affect the values of $\Delta(\rho_{1212})$ nor $\Delta(\rho_{1122})$, and resolves the problem of characterizing all quantum realizable points since now we have that the set is given by the set of all such tuples for which $r,s \in [0,1]$ and $\theta \in [0,2\pi)$. Explicitly estimating the invariants for said parametrized states yields
\begin{widetext}
\begin{equation}
    \Delta(\rho_{1122}) = \frac{1+r^2+s^2+r^2s^2+4rs\cos(\theta)}{8}
\end{equation}
and
\begin{equation}
    \Delta(\rho_{1212}) = \frac{1+r^2+s^2-r^2s^2+4rs\cos(\theta)+2r^2s^2\cos^2(\theta)}{8}.
\end{equation}
\end{widetext}
Letting $(x,y)\equiv (\Delta(\rho_{1212}),\Delta(\rho_{1122}))$, we want to show that every point parametrized as such lies inside the set $\mathfrak{B}(\mathcal{W})$ given by Eq.~\eqref{eq:quantum_set_proof_appendix}. Note that
\begin{align}
    y-x &= \frac{2r^2s^2-2r^2s^2\cos^2(\theta)}{8}\\
    &=\frac{r^2s^2}{4}\sin^2(\theta) \geq 0
\end{align}
for all $r,s,\theta$ thus implying that $x\leq y \leq 1$. Now, to show that $x \geq y^2$ we'll show that $64 (x-y^2) \geq 0$. Calculating this term explicitly, since
\begin{equation}
    y^2 = \frac{1}{64}\left((1+r^2)(1+s^2)+4rs\cos(\theta)\right)^2,
\end{equation}
we get, after expanding over all terms, 
\begin{align*}
    64 (x-y^2) &=  7 + 6r^2 - r^4 + 6s^2 - 12r^2s^2 - 2r^4s^2 \\
    &- s^4 - 2r^2s^4 - r^4s^4 + 24rs\cos(\theta) \\
    &- 8r^3s\cos(\theta) - 8rs^3\cos(\theta) - 8r^3s^3\cos(\theta).
\end{align*}
We can now re-arrange these as
\begin{align*}
    64 (x-y^2) &= 7 + 6(r^2+s^2)\\
    &-(r^4+12r^2s^2+2r^4s^2+s^4+2r^2s^4+r^4s^4)\\
    &+\cos(\theta)\, rs\bigr(24-8(r^2+s^2+r^2s^2)\bigr)\\
    &\equiv A+\cos(\theta)B.
\end{align*}
Now, since $$B = 8rs(3-r^2-s^2-r^2s^2) \geq 0$$ for all $0 \leq r,s \leq 1$, and $-1\leq \cos(\theta) \leq 1$, the minimum of $A+\cos(\theta)B$ is obtained at $\cos(\theta)=-1$. For this case, we have 
\begin{widetext}
\begin{align*}
    A-B &= (7+4rs-s^2-r^2(1+s^2))(1-4rs+s^2+r^2(1+s^2))=(7-C)(1+C)
\end{align*}
\end{widetext}
where we set 
\begin{align*}
C &= r^2(1+s^2)+s^2-4rs \\
&\geq 2rs+r^2s^2-4rs\\
&= rs(rs-2).
\end{align*}Since $0 \leq rs \leq 1$ we have that $-1\leq C\leq 3$. We conclude the argument
\begin{equation*}
    64 (x-y^2) \geq A-B = (7-C)(1+C)=7+6C-C^2\geq 0
\end{equation*}
since, for $-1\leq C\leq 7$ (and thus, in particular, for $-1\leq C\leq 3$) the function $f(C) = 7+6C-C^2 \geq 0$ as can be seen from its zeros and the sign of its second derivative.

As a final remark, we show that 
\begin{equation}
\mathfrak{Q}(\mathcal{W}) = \{\lambda \vec z \mid \lambda \geq 0, \vec z \in \mathfrak{B}(\mathcal{W})\} \equiv \mathrm{Cone}[\mathfrak{B}(\mathcal{W})].
\end{equation}
Take any unnormalized pair $\vec \rho = (\rho_1,\rho_2)$, and define $t_\ell = \mathrm{Tr}(\rho_\ell)$. If $t_\ell = 0$ for some $\ell \in \{1,2\}$ we have that $\Delta(\rho_{1,1,2,2}) = \Delta(\rho_{1,2,1,2}) = 0$, which is a point in $ \mathrm{Cone}[\mathfrak{B}(\mathcal{W})]$. If, instead, $t_{1},t_2 \neq 0$ we have that for any such pair of states we can define the re-scaled pair $\vec \rho\,' = (\rho_1/t_1,\rho_2/t_2)$. In this case, 
\begin{equation}
    \vec \Delta(\vec \rho\,) = t_1^2t_2^2 \vec \Delta(\vec \rho \,')
\end{equation}
from which we conclude that $\mathfrak{Q}(\mathcal{W}) \subseteq \mathrm{Cone}[\mathfrak{B}(\mathcal{W})]$. The converse inclusion holds trivially. 
\end{appendix}

\end{document}